\documentclass{article}
\usepackage{arxiv}
\usepackage[numbers,sort&compress]{natbib}
\usepackage[utf8]{inputenc} 
\usepackage[T1]{fontenc}    
\usepackage{ dsfont }
\usepackage{hyperref}       
\usepackage{url}
\usepackage{textcomp}
\usepackage{ gensymb }
\usepackage[ruled,vlined]{algorithm2e}
\usepackage{algorithmic}
\usepackage{booktabs}       
\usepackage{amsfonts}       
\usepackage{nicefrac}       
\usepackage{microtype}      
\usepackage{lipsum}
\usepackage{float}
\usepackage{graphicx}
\usepackage{multirow}
\usepackage{array}
\usepackage{comment}
\usepackage{longtable}
\usepackage{ragged2e}
\usepackage{subfig}

\title{Invariant Scattering Transform for Medical Imaging}

\author{
  Md Manjurul Ahsan \\
  School of Industrial and Systems Engineering\\
  University of Oklahoma\\
  Norman, Oklahoma-73019 \\
  \texttt{ahsan@ou.edu} \\
   \And
 Shivakumar Raman \\
  School of Industrial and Systems Engineering\\
  University of Oklahoma\\
  Norman, Oklahoma-73019\\
  \texttt{raman@ou.edu}\\
  \And
 Zahed Siddique \\
  School of Aerospace and Mechanical Engineering\\
  University of Oklahoma\\
  Norman, Oklahoma-73019\\
  \texttt{zsiddique@ou.edu}} 

\begin{document}
\maketitle
\begin{abstract}
Over the years, the Invariant Scattering Transform (IST) technique has become popular for medical image analysis, including using wavelet transform computation using Convolutional Neural Networks (CNN) to capture patterns' scale and orientation in the input signal. IST aims to be invariant to transformations that are common in medical images, such as translation, rotation, scaling, and deformation, used to improve the performance in medical imaging applications such as segmentation, classification, and registration, which can be integrated into machine learning algorithms for disease detection, diagnosis, and treatment planning. Additionally, combining IST with deep learning approaches has the potential to leverage their strengths and enhance medical image analysis outcomes. This study provides an overview of IST in medical imaging by considering the types of IST, their application, limitations, and potential scopes for future researchers and practitioners.

\end{abstract}
\maketitle
\section{Introduction}\label{sec1}
The Invariant Scattering Transform (IST) is a signal processing technique that was introduced by Mallat in 2012 as a way to extract high-level features from complex data that are invariant to certain transformations. IST is based on a series of wavelet transforms that capture the scale and orientation of patterns in the input signal. These wavelet transforms are computed using a Convolutional Neural Network (CNN) to process signals with spatial or temporal structures, such as images or videos~\cite{bruna2013invariant, jayalakshmy2021scattering}. 

IST is designed to be invariant to certain transformations common in real-world signals, such as translation, rotation, scaling, and deformation. This is achieved by applying a set of non-linear operators to the wavelet coefficients that generate a set of scattering coefficients. The scattering coefficients are then used to represent the signal in a way that is invariant to the selected transformations while preserving the underlying structure of the signal~\cite{simard2000transformation, sifre2013rotation}. 

In medical imaging, IST has been applied to various imaging modalities, such as chest Radiography (X-ray), Computed Tomography (CT) scan, Magnetic Resonance Imaging (MRI), and ultrasound, to extract useful features from images invariant to variations in imaging conditions~\cite{haque2020deep}. For example, IST can extract edges, contours, and texture patterns from the images, which is essential for various medical imaging tasks, such as segmentation, registration, and diagnosis~\cite{jiang2010medical}. 
IST has several advantages over other feature extraction techniques in medical imaging such as~\cite{jiao2021multi, salahat2017recent, pereira2022melanoma}:

\begin{itemize}

    \item IST is designed to be invariant to certain transformations, which can improve the robustness and accuracy of feature extraction.
    \item IST can capture multi-scale and multi-resolution features, improving the features' discrimination and interpretability.
    \item IST is computationally efficient and easily integrated into medical imaging pipelines.
\end{itemize}

Therefore, IST is a powerful signal-processing technique that can be applied to medical imaging to extract valuable features invariant to certain transformations. The application of IST to medical imaging is an emerging area of research that has the potential to advance the field and contribute to the development of new clinical tools and applications.

The use of IST in medical imaging has the potential to offer several benefits, including~\cite{souli2021robust, liu2020wavelet, bruna2022scattering, martinez2022comparison}:
\begin{itemize}

    \item \textbf{Robust Feature Extraction:} IST is a non-parametric feature extraction technique robust to image variations, such as translation, rotation, scaling, and deformation. This makes IST a valuable tool for medical image analysis, where images often exhibit significant variability due to differences in anatomy, acquisition, and patient positioning.
    \item \textbf{Computational Efficiency:} IST is computationally efficient and well-suited for large-scale medical imaging applications.
    \item \textbf{Multiscale Representation:} IST provides a multiscale representation of images that captures local and global image features. This can improve the performance of medical imaging algorithms by providing a more comprehensive and informative representation of images.
    \item \textbf{Visualization:} IST can be used to create visual representations of images that highlight the most important features of a given task. This can be useful for medical diagnosis, as it enables doctors to see which areas of an image are most relevant to a particular diagnosis.
    \item \textbf{Explainable AI:} IST can be used to create explainable AI models that provide post-hoc explanations of how a model arrived at its decision. This is important in medical imaging, as it enables doctors to understand why a model made a particular diagnosis, which can enhance the transparency and accountability of medical imaging algorithms.
\end{itemize}
Therefore, applying IST can improve medical imaging algorithms' accuracy, efficiency, and transparency, leading to better medical diagnosis and improved patient outcomes.

This study aims to provide a comprehensive overview of the different types of IST, their relation to signal processing, and their applications, limitations, and future directions in medical imaging. The remaining sections are organized as follows: Section Two provides relevant background and basic information regarding IST. Section Three evaluates some of the reference literature that has considered IST in medical imaging.  Section Four discusses the overall findings and suggests future research directions. Finally, in Section Six, a general conclusion is drawn based on the overall analysis of IST in medical imaging.

\section{IST Background}
This section provides a general background and relevant information on various aspects of the IST in medical imaging. Specifically, the type of IST used, signal processing techniques, deep learning approaches, and the types of datasets utilized for the experiments are discussed in detail to offer a comprehensive overview of IST in medical imaging.

As mentioned earlier, the IST is a multi-scale transform that provides a detailed analysis of signals and images. This transform is based on wavelets and invariant to certain input signal transformations. By capturing information about the structure of the signal at different scales, IST provides a powerful tool for analyzing and processing signals and images. There are several types of IST, each of which provides a different perspective on the input signal and can be used in different applications. Understanding the different types of IST and their applications is crucial for effectively using this transform in signal and image processing. Among the various types of IST, there are three major types that are widely used by researchers. These include~\cite{mallat2012group, sifre2012combined, curtright2023scale}:
\begin{itemize}
    \item Translation-Invariant Scattering (TIS)
    \item Rotation-Invariant Scattering (RIS)
    \item Scale-Invariant Scattering (SIS)
\end{itemize}
\begin{enumerate}
    \item \textbf{TIS} is an IST invariant to the translation of the input signal. The transform is based on wavelets and provides a multi-scale signal representation, capturing information about its local structure. The following equation gives the TIS transform:
    \begin{equation}
S_{j,k} = \sum_{m \in \mathbb{Z}} \left\vert T_{j, k + 2^jm} \ast f \right\vert^2
\end{equation}
where $f$ is the input signal, $T_{j, k}$ are wavelets at scale $j$ and position $k$, and $\ast$ represents convolution. This transform provides information about the local features of the signal, such as edges and textures, and is widely used in image and signal processing applications~\cite{mallat2012group}.
\item \textbf{RIS} is a type of IST that is invariant to rotations of the input signal. The transform is based on wavelets and provides a multi-scale signal representation, capturing information about its symmetries. The following equation gives the RIS transform:
\begin{equation}
S_{j,k} = \sum_{m = 0}^{2^{j} - 1} \left\vert T_{j, k + m} \ast f \right\vert^2
\end{equation}

where $f$ is the input signal, $T_{j, k}$ are wavelets at scale $j$ and position $k$, and $\ast$ represents convolution. This transform provides information about the symmetries of the signal, such as rotations and reflections, and is widely used in image and signal processing applications~\cite{sifre2013rotation}.

Figure~\ref{fig:rotate}, presents an illustrative depiction of the principles of rotation co- and invariance in image processing, as mentioned in~\cite{bekkers2018roto}. Specifically, the image portrays a star-shaped object rotated at different angles, showcasing the distinction between rotational covariance and invariance. The left half of the image exemplifies rotational covariance, where the features of the star change with rotation, resulting in different feature representations. Conversely, the right half of the image showcases the concept of rotational invariance, where the star's features remain unaltered regardless of the rotation angle, leading to identical feature representations. This visual representation provides a lucid and succinct demonstration of these fundamental concepts and their crucial significance in image processing and computer vision applications.
\begin{figure}[H]
    \centering
    \includegraphics[width=\textwidth]{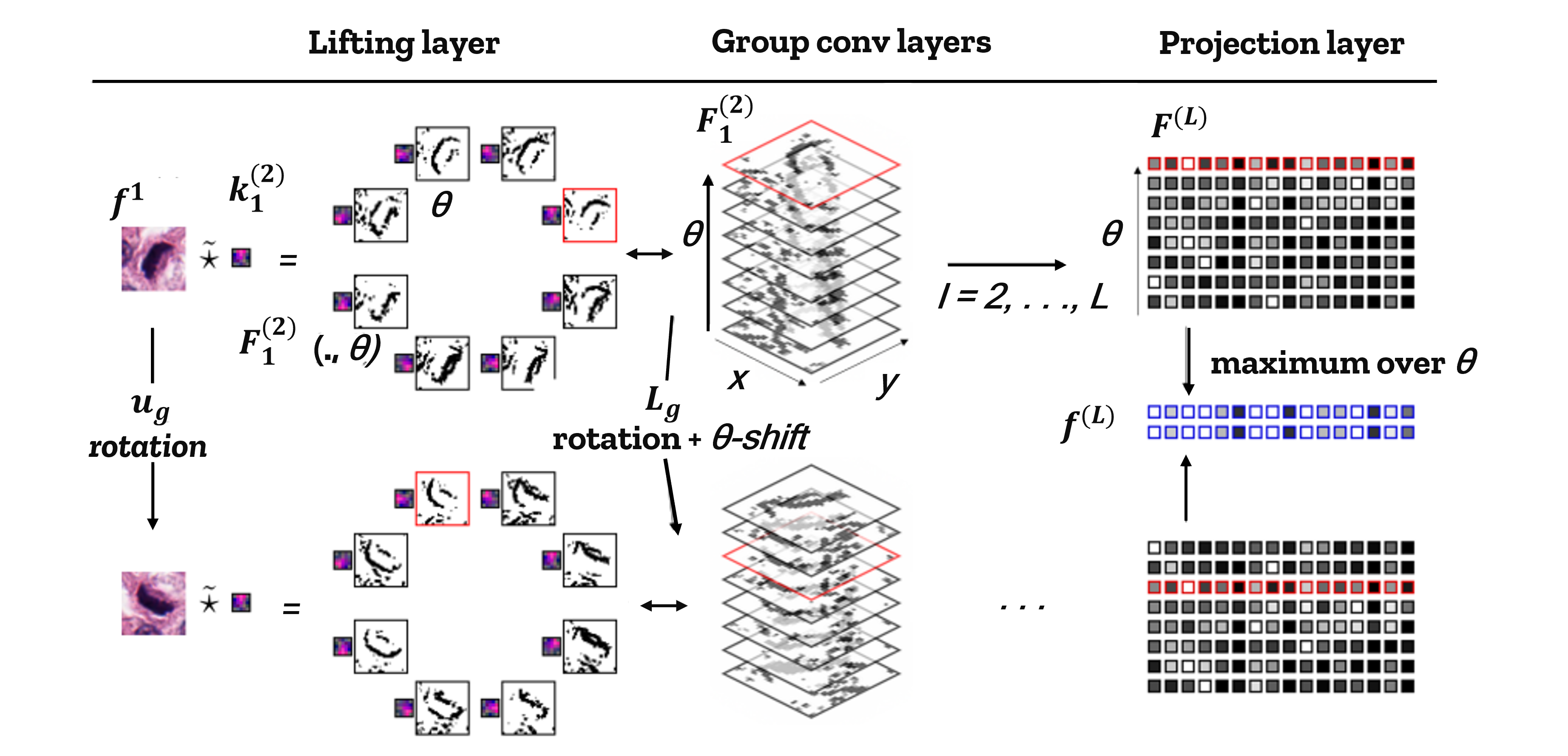}
    \caption{Illustration of rotation co- and invariance retrieved from~\cite{bekkers2018roto}}
    \label{fig:rotate}
\end{figure}
\item \textbf{SIS} is a type of IST invariant to the input signal's scalings. The transform is based on wavelets and provides a multi-scale signal representation, capturing information about its global structure. The following equation gives the Scale-Invariant Scattering transform:
\begin{equation}
S_{j,k} = \sum_{m = 0}^{j} \left\vert T_{j - m, k} \ast f \right\vert^2
\end{equation}

where $f$ is the input signal, $T_{j, k}$ are wavelets at scale $j$ and position $k$, and $\ast$ represents convolution. This transform provides information about the global structure of the signal, such as its overall shape and size, and is widely used in image and signal processing applications.

\end{enumerate}

\begin{figure}[H]
    \centering
    \includegraphics[width=\textwidth]{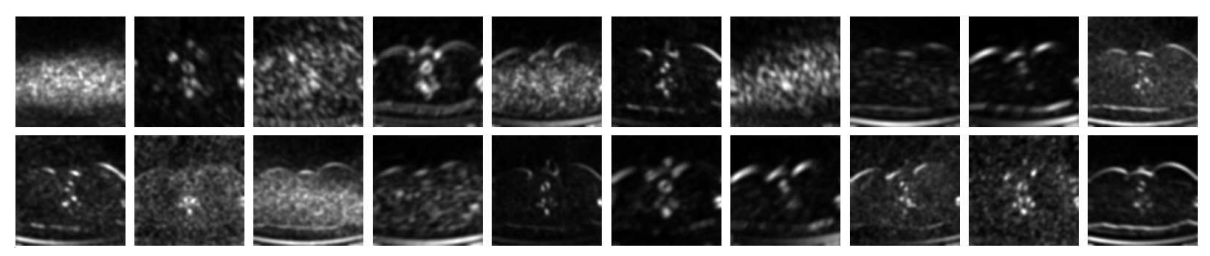}
    \caption{visual depictions of select scattering transform outcomes derived from the initial computed tomography (CT) image from~\cite{lan2016medical} .}
    \label{fig:sct}
\end{figure}

Figure \ref{fig:sct} presents a collection of visual representations showcasing selected scattering transform outcomes obtained from the primary CT image, as featured in~\cite{lan2016medical}. The depicted images illustrate the results of the scattering transform applied to the original CT image and demonstrate the efficacy of this method in capturing and enhancing various features in the input data. These visualizations provide an intuitive and informative overview of the scattering transform's ability to extract meaningful information from complex medical imaging data.

Table~\ref{tab:tabb1} compares TIS, RIS, and SIS regarding their underlying type of invariance, the information captured about the input signal, and their applications. The table lists each type of IST in the first column, the type of invariance in the second column, the captured information in the third column, and some common applications in the fourth column. Therefore, the comparison presented in Table~\ref{tab:tabb1} offers a thorough understanding of the distinctions between the three types of IST and highlights their respective advantages and limitations.
\begin{table}[H]
\caption{Different Types of IST and Their Applications}
\label{tab:tabb1}
\resizebox{\columnwidth}{!}{%
\begin{tabular}{@{}llll@{}}
\toprule
\textbf{Transform} & \textbf{Invariance} & \textbf{Captures} & \textbf{Applications} \\ \midrule
\begin{tabular}[c]{@{}l@{}}Translation-\\ Invariant \\ Scattering\end{tabular} & Translations & \begin{tabular}[c]{@{}l@{}}Local \\ structure\end{tabular} & \begin{tabular}[c]{@{}l@{}}Image \\ segmentation, Image \\ denoising, Image \\ registration, Image classification\end{tabular} \\
\begin{tabular}[c]{@{}l@{}}Rotation-\\ Invariant \\ Scattering\end{tabular} & Rotations & Symmetries & \begin{tabular}[c]{@{}l@{}}Image analysis \\ in polar coordinates, Analysis of symmetrical \\ structures in medical images, such as the human \\ brain or heart, Detection of rotational \\ symmetries in \\ medical images, such as in the case of certain\\ diseases or conditions\end{tabular} \\
\begin{tabular}[c]{@{}l@{}}Scale-\\ Invariant \\ Scattering\end{tabular} & Scalings & \begin{tabular}[c]{@{}l@{}}Global \\ structure\end{tabular} & \begin{tabular}[c]{@{}l@{}}Analysis of the overall structure of medical images, \\ such as the size and shape of \\ organs or tissues, Detection of changes in the size \\ and shape of structures \\ over time,\\ Segmentation of multi-scale structures \\ in medical images, \\ such as blood \\ vessels or tumors\end{tabular} \\ \bottomrule
\end{tabular}%
}
\end{table}

\begin{figure}[H]
    \centering
    \includegraphics[width=\textwidth]{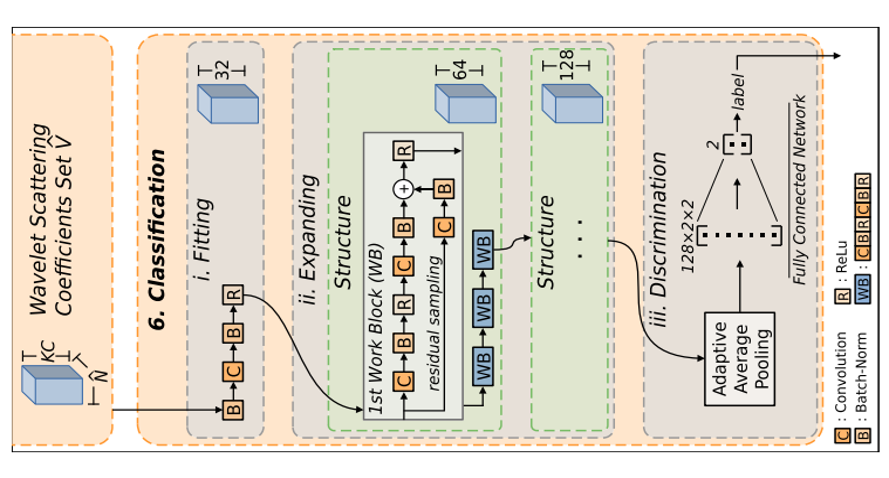}
    \caption{Illustration of a proposed model pipeline by Pereira et al. (2022) consisting of the scattering transform and Convolutional Neural Network (CNN)~\cite{pereira2022melanoma}.}
    \label{fig:IST1}
\end{figure}

Figure \ref{fig:IST1} illustrates the proposed model pipeline by Pereira et al. (2022) that consists of the scattering transform and CNN for the effective retrieval of medical images. The proposed approach uses a feature-set Vˆ obtained from the scattering coefficients to train a deep learning model comprising three primary components. These include an initial data fitting stage, a primary processing component that employs convolutions to expand the given data volume, and a fully connected layer. The results of the study suggest that the proposed model pipeline has the potential to achieve higher accuracy and precision rates in the retrieval of medical images~\cite{pereira2022melanoma}.

The IST is an integral component of the popular signal processing technique that leverages both convolution and wavelet transform to achieve invariance and stability. The IST employs a hierarchical methodology that dissects signals into a sequence of wavelet transform coefficients and subsequently utilizes a series of invariant scattering operations to generate a robust representation of the signal. The key principles and concepts of signal processing that are germane to the IST comprise the mathematical operation of convolution, the mathematical tool of Wavelet Transform, and the properties of invariance and stability, which are described below: 
\subsection{Signal Processing}
\begin{itemize}

    \item \textbf{Convolution:} A mathematical operation that combines two functions to produce a third, reflecting how one function modifies the other. In IST, convolution is used to apply filters or kernels to the input image or signal to extract relevant features. The formula of the convolution can be expressed as follows~\cite{albawi2017understanding}: 
    \begin{equation}
(f * g)(t) = \int_{-\infty}^{\infty} f(\tau)g(t-\tau) d\tau
\end{equation}

In this formula, $f$ and $g$ are two functions, and the convolution of $f$ and $g$ is denoted by $(f * g)(t)$. The integral is taken over the entire real line, with $f$ evaluated at $\tau$ and $g$ evaluated at $t-\tau$. The result is a new function of $t$, which represents the convolution of $f$ and $g$ at time $t$.
    \item \textbf{Wavelet Transform:} A mathematical tool that decomposes a signal or image into components at different scales and frequencies. IST leverages wavelet transforms to provide multi-scale representations of the input data, capturing both local and global structures. The formula of the Wavelet Transform can be expressed as follows~\cite{jensen2001ripples}:  
    \begin{equation}
W_{f}(a, b) = \frac{1}{\sqrt{a}}\int_{-\infty}^{\infty} f(t)\psi\left(\frac{t-b}{a}\right) dt
\end{equation}

In this formula, $f(t)$ is the function we want to analyze, $\psi(t)$ is the wavelet function, and $a$ and $b$ are parameters that control the scale and translation of the wavelet. The wavelet transform at scale $a$ and position $b$ is denoted by $W_{f}(a, b)$.

The integral is taken over the entire real line, with $f$ evaluated at $t$ and $\psi$ evaluated at $(t-b)/a$. The result is a new function of $a$ and $b$, which represents the wavelet transform of $f$ at the given scale and position. The factor of $1/\sqrt{a}$ ensures that the wavelet coefficients are appropriately scaled for different values of $a$.
    \item \textbf{Invariance:} In signal processing, invariance refers to the property of a system that remains unchanged under certain transformations, such as translation, rotation, or scaling. IST aims to provide invariant features to improve the robustness and generalization of the analysis. The formula of the Invariance can be expressed as follows~\cite{sifre2013rotation}:  
    \begin{equation}
T(g(x)) = T(x)
\end{equation}

In this formula, $T$ is an operator that represents a certain kind of invariance, and $x$ is an input to that operator. $g(x)$ is a transformation applied to $x$, and the formula expresses the idea that if $g(x)$ is applied to $x$, the resulting output $T(g(x))$ should be the same as the output of the operator applied to the original input $x$, i.e. $T(x)$.

This kind of invariance is important in many fields, including computer vision and machine learning, where it is often desirable for models to be able to recognize the same object or pattern regardless of its orientation, scale, or other variations. By designing operators that are invariant to certain kinds of transformations, it is possible to build more robust and accurate models that can generalize better to new inputs.

    \item \textbf{Hierarchical Feature Extraction:} Deep learning models, such as CNN, learn hierarchical representations of the input data by applying multiple layers of convolution and pooling. Similarly, IST extracts feature hierarchically, with each layer capturing increasingly complex structures~\cite{jogin2018feature}. 
    \item \textbf{Non-linear Activation Functions:} Deep learning models often use non-linear activation functions, like ReLU or sigmoid, to introduce non-linearity and improve the expressive power of the model~\cite{ahsan2021detecting}. IST also incorporates non-linearities in its processing pipeline to capture complex relationships in the data.
    \item \textbf{Pooling:} Pooling operations, such as max-pooling or average-pooling, are used in deep learning to reduce the spatial dimensions of feature maps and achieve a degree of invariance~\cite{gholamalinezhad2020pooling}. IST applies similar concepts to aggregate information across different scales and orientations, contributing to its invariance properties. 
    \item \textbf{End-to-End Learning:} Deep learning models learn to extract features and make predictions simultaneously in an end-to-end fashion. Although IST is not a fully end-to-end learning method, its features can be incorporated into deep learning models to improve performance and generalization~\cite{xu2017end}. 
\end{itemize}
In summary, IST leverages key concepts and principles from signal processing, such as convolution, wavelet transform, invariance, stability, and deep learning, including hierarchical feature extraction, non-linear activation functions, pooling, and end-to-end learning. These concepts contribute to IST's robustness, invariance, and performance in various medical imaging applications.
\subsection{Challenges in Medical Image Processing and IST Solutions}

Existing techniques for processing medical images have several limitations, which the IST aims to address. Below, we discuss some of these potential challenges and how IST can help overcome them~\cite{javadpour2016improving, reena2018improved, curtright2023scale, bruna2013invariant}: 
\begin{itemize}

    \item \textbf{Sensitivity to Noise and Artifacts:} Traditional image processing techniques can be sensitive to noise and artifacts in medical images. The IST provides more robust features by incorporating wavelet transforms and hierarchical processing, which can effectively handle noise and maintain stability.

    \item \textbf{Lack of Invariance:} Some existing techniques may not provide invariant features, which are crucial for medical imaging since anatomical structures can appear in different positions, orientations, and scales. IST focuses on extracting features that are invariant to transformations like translation, rotation, and scaling, improving the generalization and performance of the analysis. 
    \item \textbf{Complexity in Feature Engineering:} Conventional methods often rely on handcrafted features, which can be complex and time-consuming to design. IST uses a hierarchical approach to feature extraction, automatically learning informative and invariant features without the need for extensive manual feature engineering.
    \item \textbf{Limited Adaptability:} Traditional techniques may be designed for specific imaging modalities, requiring separate methods for different types of medical images. IST is a versatile framework that can be adapted to various imaging modalities, offering a unified approach to feature extraction and analysis.
    \item \textbf{Scalability and Computational Efficiency:} Some existing methods may not be computationally efficient, hindering their application to large-scale datasets and real-time processing tasks. IST is designed with computational efficiency in mind, making it suitable for large-scale medical image analysis and real-time applications. 
    \item \textbf{Integration with Machine Learning Algorithms:} Traditional image processing techniques may not be easily integrated with machine learning algorithms, limiting their potential in advanced medical image analysis tasks. The features extracted by IST can be readily integrated into various machine learning models, including deep learning architectures, to build powerful and effective systems for medical image analysis.
\end{itemize}
While IST addresses many limitations of existing techniques, it is worth noting that it may have its own limitations, such as the need for parameter tuning and potential challenges in interpretability. Nevertheless, IST offers a promising framework for medical image processing by providing robust, invariant, and informative features that can be integrated into machine learning algorithms to enhance performance and generalization. 
\subsection{Key Steps and Process to Apply IST in Medical Imaging}

Applying the IST to medical images involves several key steps, which can be summarized as follows: 
\begin{itemize} 
    \item \textbf{Preprocessing:} Before applying IST, medical images may require preprocessing to enhance their quality and ensure compatibility with the IST framework. This may include tasks such as noise reduction, artifact removal, image normalization, and resizing.

Figure~\ref{fig:pre} provides a visual representation of the preprocessing procedure flowchart for the Monkeypox image datasets, illustrating the steps involved in resizing the images using zero-padding. The reference for this information and the accompanying image is retrieved from the article by Ahsan et al. (2023)~\cite{ahsan2023deep}.
    \begin{figure}
        \centering
        \includegraphics[width=\textwidth]{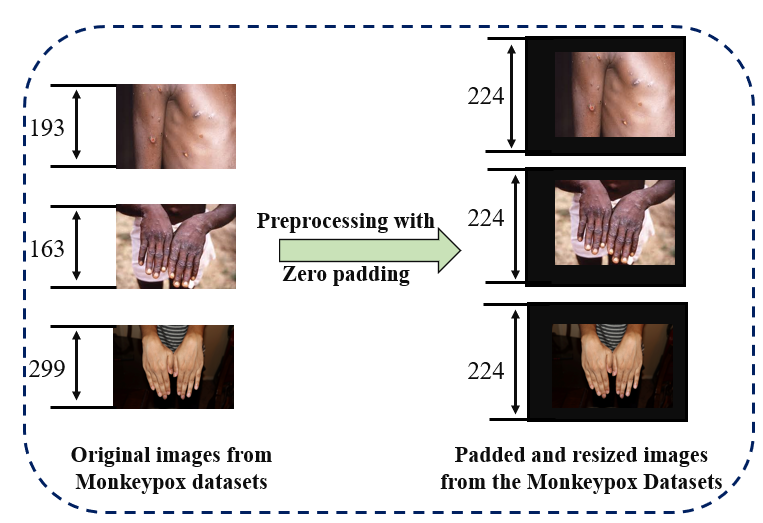}
        \caption{Illustration of preprocessing with zero padding retrieved from~\cite{ahsan2023deep}.}
        \label{fig:pre}
    \end{figure}
    \item \textbf{Construction of Wavelet Filters:} IST relies on wavelet transforms to analyze images at different scales and frequencies. The construction of wavelet filters, such as Gabor or Morlet wavelets, is a crucial step, as these filters will be applied to the input image to capture its underlying structures.

    Figure~\ref{fig:gabor} illustrates the convolution kernels of 2D Gabor filters obtained from the reference by Hu et al. (2020). The image depicts the different Gabor filters and their orientations, scales, and spatial frequencies. Gabor filters are used in image processing and computer vision tasks for feature extraction, edge detection, and texture analysis~\cite{hu2020gabor}. The figure's visualization of the Gabor filters highlights their ability to capture different image features by convolving them with the input image.
    \begin{figure}
        \centering
        \includegraphics[width=\textwidth]{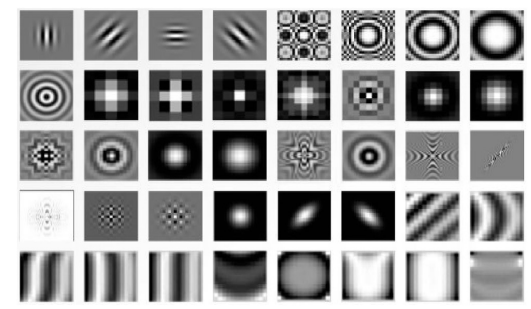}
        \caption{Illustration of convolution kernels of 2D Gabor filters retrieved from~\cite{hu2020gabor}.}
        \label{fig:gabor}
    \end{figure}
    \item \textbf{Convolution with Wavelet Filters:} Apply the constructed wavelet filters to the input image using the convolution operation. This step results in a set of wavelet coefficients that represent the local frequency content of the image at different scales and orientations~\cite{liu2019multi}. 
    \item \textbf{Non-Linear Modulation:} Apply a pointwise non-linear modulation, such as taking the absolute value or the square of the wavelet coefficients. This step introduces non-linearity into the IST framework, improving its expressive power and stability~\cite{liu2018optimization}.
    \item \textbf{Averaging and Pooling:} To achieve invariance to certain transformations, such as translation and rotation, perform an averaging or pooling operation on the modulated wavelet coefficients. This step reduces the spatial dimensions and combines information across different scales and orientations~\cite{gholamalinezhad2020pooling}. 

\item \textbf{Iterative Hierarchical Processing:} Repeat steps 3 to 5 in a hierarchical manner, applying wavelet transforms, non-linear modulation, and pooling operations iteratively on the output from the previous layer. This creates a multi-layered representation of the input image, with each layer capturing increasingly complex structures~\cite{martins2016novel}. 

\item \textbf{Feature Extraction:} After the hierarchical processing is complete, extract the final set of features from the output of the last layer. These features should be invariant and informative, representing the essential characteristics of the input medical image~\cite{guyon2008feature}. 

\item \textbf{Integration with Machine Learning Algorithms:} The extracted features can be used as input to various machine learning algorithms, such as support vector machines, random forests, or deep learning models, for tasks like classification, segmentation, or detection. 

The flow diagram of a deep learning-based model that incorporates extracted features for COVID-19 diagnosis is demonstrated in Figure~\ref{fig:dlmodel}. The image was retrieved from reference~\cite{ahsan2020covid}, which provides further details on the development and evaluation of the model. The flow diagram illustrates the different components of the model, including the pre-processing of input data, feature extraction, and classification. The use of deep learning-based models for COVID-19 diagnosis has shown promise in recent studies, and the diagram in Figure~\ref{fig:dlmodel} provides a useful visual representation of the model architecture.
\begin{figure}
    \centering
    \includegraphics[width=\textwidth]{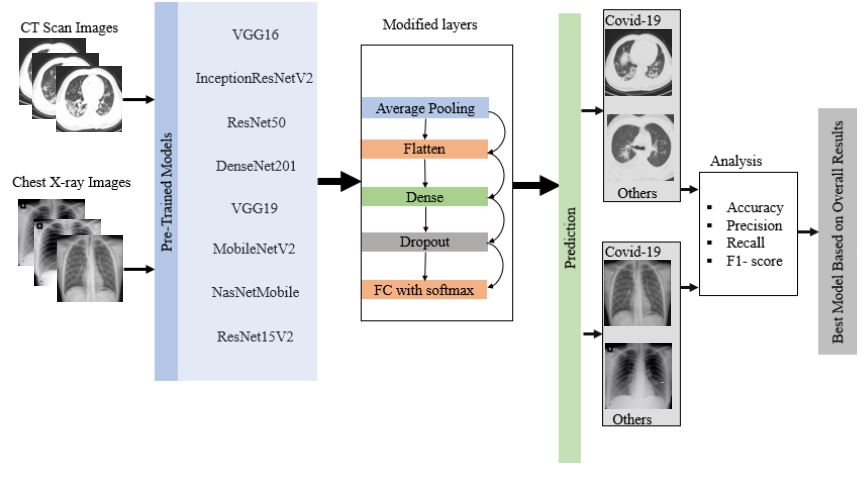}
    \caption{Demonstration of the flow diagram of deep learning-based model incorporated with extracted features for COVID-19 diagnosis retrieved from~\cite{ahsan2020covid}.}
    \label{fig:dlmodel}
\end{figure}

\item \textbf{Model Training and Evaluation:} Train the chosen machine learning model using the extracted IST features and a labeled dataset of medical images. Evaluate the performance of the model using appropriate metrics, such as accuracy, precision, recall, or the F1-score~\cite{bi2019machine}.

\end{itemize}
In summary, applying IST to medical images involves preprocessing, constructing wavelet filters, convolving the image with these filters, applying non-linear modulation, averaging and pooling, and iterative hierarchical processing. The extracted features can then be integrated into machine-learning algorithms for various medical image analysis tasks.
\subsection{IST Parameters and Settings: Impact on Performance}
The IST has several key parameters and settings that influence its performance in feature extraction and analysis. The choice of these parameters affects the quality, invariance, and computational efficiency of the extracted features. Some of the key parameters and settings include~\cite{mallat1999wavelet, ahsan2020deep, masci2011stacked}:

\begin{itemize}

\item \textbf{Wavelet Filters:} The choice of wavelet filters is crucial, as they capture the frequency content of the input image at different scales and orientations. Different wavelet functions, such as Gabor, Morlet, or Haar wavelets, have distinct properties that can affect the performance of IST. Selecting appropriate wavelet filters that capture the relevant information in medical images is essential for optimal results. 

\item \textbf{Filter Scales and Orientations:} The number of scales and orientations used in the wavelet filters influences the level of detail captured by IST. Using more scales and orientations can provide a richer representation of the input image but may increase the computational complexity. Balancing the trade-off between capturing sufficient information and maintaining computational efficiency is essential. 

\item \textbf{Non-linear modulation function:} The choice of the non-linear modulation function, such as taking the absolute value or the square of the wavelet coefficients, affects the stability and expressive power of IST. Selecting an appropriate non-linear function is important for capturing complex relationships in the input data. 

\item \textbf{Averaging or pooling function:} The choice of averaging or pooling function, such as max-pooling or average-pooling, affects the degree of invariance achieved by IST. Different pooling functions can lead to varying levels of invariance and performance in downstream tasks. The choice of pooling function should be guided by the specific requirements of the medical imaging task at hand.

Figure~\ref{fig:mlp-cnn} displays the flow diagram of the Multilayer Perceptron and Convolutional Neural Network (MLP-CNN) based models proposed by the authors in reference~\cite{ahsan2020deep}. The proposed models leverage both average pooling and max pooling techniques to develop an optimal framework for identifying patients with COVID-19 symptoms based on chest X-ray images. The flow diagram provides a comprehensive overview of the various stages involved in the proposed model architecture, including the preprocessing of input data, feature extraction, and classification. The use of MLP-CNN models has demonstrated promising results in various studies related to medical image analysis and diagnosis, and the model proposed in reference~\cite{ahsan2020deep} is a valuable contribution to the field of COVID-19 diagnosis.
\begin{figure}
    \centering
    \includegraphics[width=\textwidth]{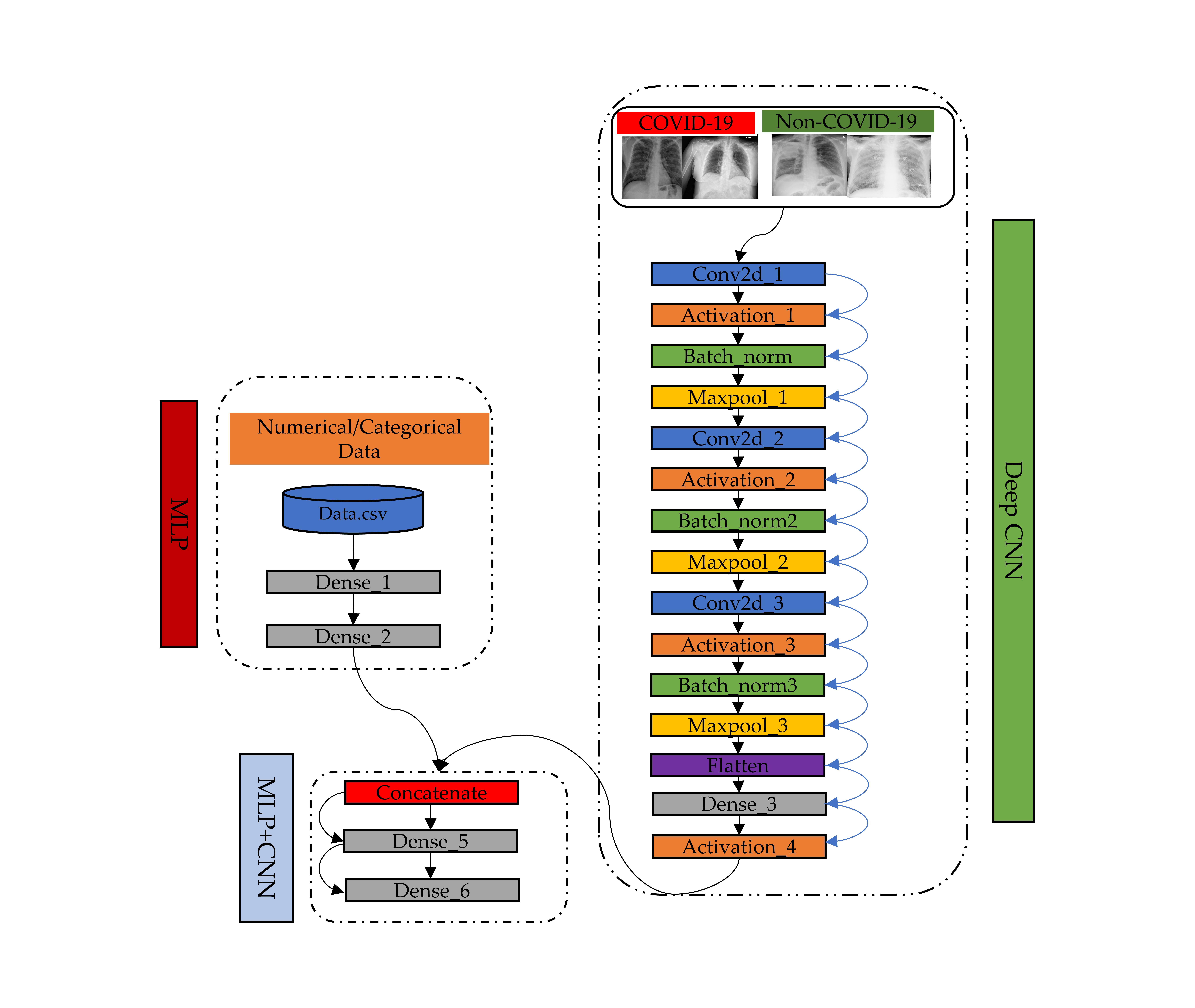}
    \caption{Flow diagram of  Multilayer Perceptron and Convolutional Neural Network (MLP-CNN) based models proposed by~\cite{ahsan2020deep}}
    \label{fig:mlp-cnn}
\end{figure}
\item \textbf{Number of Hierarchical Layers:} The number of hierarchical layers in the IST affects the complexity of the extracted features. More layers can capture increasingly complex structures, but may also increase the computational complexity and risk overfitting. Choosing an appropriate number of layers is crucial for balancing feature expressiveness and computational efficiency. 

\item \textbf{Parameter Tuning:} IST may require tuning of parameters, such as filter bandwidths, the non-linear modulation function, or the pooling function, to optimize performance for a specific task or dataset. Proper tuning of these parameters can lead to more informative and invariant features, improving the performance of downstream tasks. 
\end{itemize}
The performance of IST is highly dependent on the choice of these key parameters and settings. Selecting appropriate wavelet filters, scales, orientations, non-linear modulation functions, pooling functions, and hierarchical layers can significantly affect the quality, invariance, and computational efficiency of the extracted features. Proper tuning and selection of these parameters are essential for optimal performance in medical imaging tasks~\cite{tajbakhsh2016convolutional}. 
 \subsection{IST for Medical Image Segmentation, Classification, and Registration}
 The IST has shown promising results when applied to medical images in various contexts, including segmentation, classification, and registration. Although specific results may depend on the particular study, dataset, and imaging modality, the following provides a general overview of the outcomes observed in these applications~\cite{ji2019invariant, ahsan2022transfer}:
 \begin{itemize}
     \item 
 
 \textbf{Segmentation:} IST has been found to improve the segmentation of medical images by providing robust and informative features that capture the structural and textural information of different regions. When integrated with segmentation algorithms, IST features can help accurately delineate the boundaries of various anatomical structures, such as organs, vessels, or tumors. This improvement can potentially lead to better diagnosis, treatment planning, and image-guided interventions. 

 Figure~\ref{fig:seg} is a sample example that provides the consequences of threshold-based segmentation when applied to images of hands infected with the Monkeypox virus, as presented in~\cite{ahsan2022transfer}. The region of interest (ROI) is defined using the pixel values corresponding to the infected areas, following which a mask is generated to facilitate segmentation and produce the ultimate outcome.
 \begin{figure}
     \centering
     \includegraphics[width=.7\textwidth]{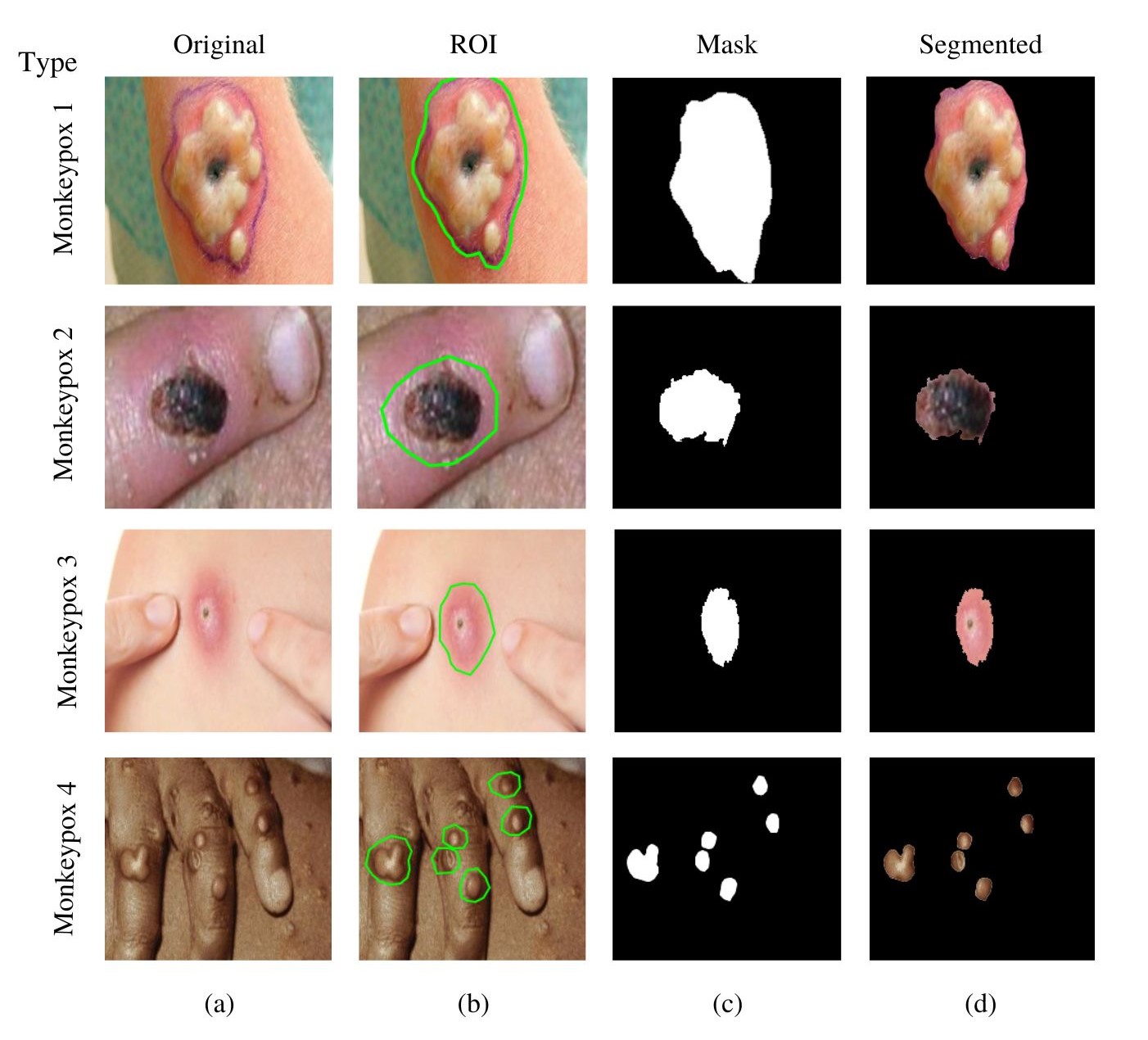}
     \caption{Demonstration of Threshold-based segmentation (a) Original, (b) ROI, (c) Mask, (d) Segmented retrieved from~\cite{ahsan2022transfer}.}
     \label{fig:seg}
 \end{figure}

\item \textbf{Classification:} IST has demonstrated its potential in medical image classification tasks, such as disease detection and diagnosis. By extracting informative and invariant features that distinguish between healthy and pathological tissues, IST can enhance the performance of machine learning algorithms in classifying images based on the presence or absence of specific diseases. Improved classification performance can contribute to more accurate and timely disease detection and diagnosis~\cite{ahsan2022machine}. 

\item \textbf{Registration:} IST features that are invariant to translation, rotation, and scaling can be beneficial for image registration tasks. By matching IST features between different images, it becomes possible to align and register images from different modalities, patients, or time points. This alignment can facilitate more accurate analysis and comparison of medical images, which is essential for tasks like longitudinal monitoring of disease progression or treatment response evaluation~\cite{stefano2021robustness}.

\end{itemize}
The application of IST to medical images in various contexts has demonstrated its ability to enhance the performance of segmentation, classification, and registration tasks. By providing robust, informative, and invariant features, IST can improve the outcomes of medical image analysis and contribute to better diagnosis, treatment planning, and monitoring. It is important to note, however, that the specific results may vary depending on the study, dataset, imaging modality, and the choice of parameters and settings for IST. 
\subsection{Dataset}
\begin{figure}[h]
    \centering
    \includegraphics[width=\textwidth]{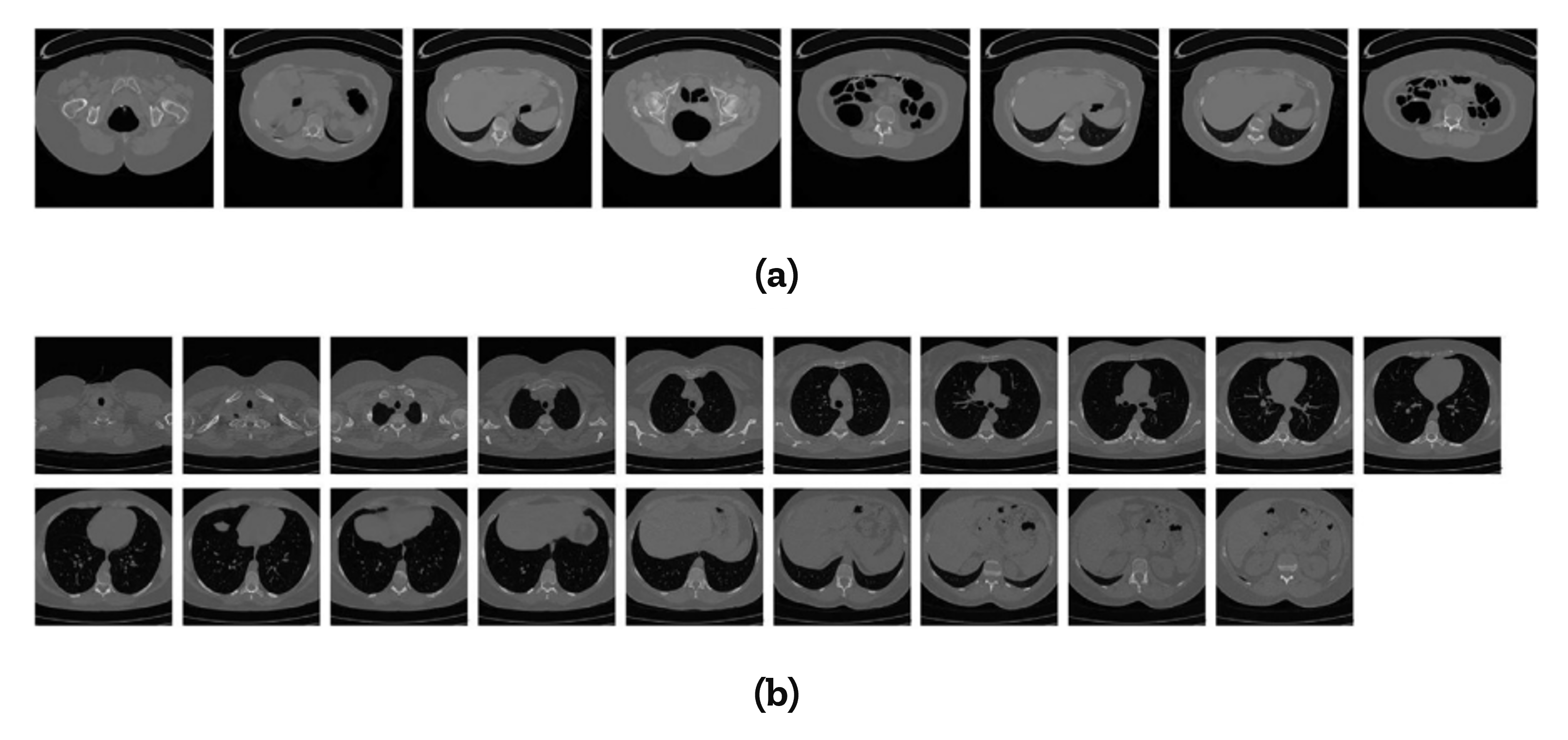}
    \caption{Representative dataset of (a) TCIA-CT and (b) EXACT09-CT datasets used in~\cite{lan2018integrated}.}
    \label{fig:dataset1}
\end{figure}

There are several open benchmark image dataset that the researchers have widely used for IST in developing ML-based diagnosis model, and some of them are described below: 
\begin{enumerate}
    \item \textbf{ISIC (International Skin Imaging Collaboration) dataset:} The ISIC dataset is a large collection of skin lesion images used for melanoma classification. It contains over 23,000 images of skin lesions, including benign and malignant melanomas, seborrheic keratoses, and nevi. The images are captured using a variety of imaging modalities, such as clinical photography, dermoscopy, and confocal microscopy, and are accompanied by diagnostic annotations provided by dermatologists. The dataset is used for training and testing machine learning algorithms for automated skin lesion diagnosis. The ISIC dataset has been widely used in the research community and has led to the development of several automated skin lesion diagnosis systems, which have shown promising results in detecting melanoma, a deadly form of skin cancer. The ISIC dataset is freely available to researchers and is continuously updated with new images and annotations to improve its utility as a research tool~\cite{gutman2016skin}.
\item \textbf{LIDC-IDRI (Lung Image Database Consortium and Image Database Resource Initiative)} The LIDC-IDRI is a publicly available computed tomography (CT) lung scan dataset. The dataset contains over 1,000 CT scans of the chest, each of which has been annotated by at least four experienced radiologists. The annotations include the location and size of nodules and an assessment of their likelihood of malignancy. The LIDC-IDRI dataset is widely used in developing ML algorithms for detecting and diagnosing lung cancer, one of the leading causes of cancer-related deaths worldwide. The dataset has facilitated the development of several automated lung nodule detection and classification systems, which have shown promising results in detecting early-stage lung cancer. The LIDC-IDRI dataset is freely available to researchers and has led to significant advances in medical imaging for lung cancer diagnosis and treatment~\cite{lin2018lung}.
\item \textbf{BraTS (Multimodal Brain Tumor Segmentation Challenge)} The BraTS dataset is widely used in the field of medical imaging. It consists of magnetic resonance imaging (MRI) scans of the brain that have been annotated for the presence and location of brain tumors. The dataset contains images from adult and pediatric patients, including four types of MRI scans: T1, T1-contrast-enhanced, T2, and T2-FLAIR. The annotations provided with the dataset include segmentation masks for the tumor core, the enhancing tumor, and the whole tumor. The BraTS dataset has been used to evaluate the performance of various ML algorithms for automated brain tumor segmentation and diagnosis. The dataset has facilitated the development of several automated brain tumor segmentation systems, which have shown promising results in detecting and characterizing brain tumors. The BraTS dataset is freely available to researchers and has contributed significantly to developing new imaging techniques and ML algorithms for brain tumor diagnosis and treatment~\cite{menze2012multimodal}.
\item \textbf{CAMELYON16 (The Cancer Metastases in Lymph Nodes Challenge 2016) dataset} The Cancer Metastases in Lymph Nodes Challenge 2016 (CAMELYON16) dataset is a publicly available dataset of histopathology images used to detect breast cancer metastases in lymph nodes. The dataset contains over 400 digital whole-slide images of hematoxylin and eosin (H\&E) stained lymph node sections acquired from two different centers, with 2,792 regions of interest (ROIs) annotated by pathologists. The ROIs contain either metastatic or normal tissue, and the goal is to develop automated algorithms that can accurately identify the metastatic regions. The dataset has been widely used to evaluate the performance of ML algorithms for automated lymph node metastasis detection. The CAMELYON16 dataset has facilitated the development of several automated lymph node metastasis detection systems, which have shown promising results in detecting cancerous regions with high accuracy~\cite{chen2016identifying}. 
\item \textbf{MURA (Musculoskeletal Radiographs) dataset} The MURA dataset is a publicly available dataset of radiographic images used to detect abnormal musculoskeletal conditions. The dataset contains over 40,000 musculoskeletal radiographs, including shoulder, elbow, wrist, hand, hip, knee, ankle, and foot images. The images are labeled as normal or abnormal, and abnormal images are classified into one of seven categories: arthritis, fracture, ligament tear, dislocation, osteoporosis, joint effusion, or soft tissue injury. The MURA dataset has been widely used to evaluate the performance of machine learning algorithms for the automated detection of abnormal musculoskeletal conditions. The dataset has facilitated the development of several automated musculoskeletal image analysis systems, which have shown promising results in detecting and diagnosing various musculoskeletal conditions~\cite{rajpurkar2017mura}. 

A sample dataset incorporated with IST used in medical image retrieval by Kashem et al. (2022) is presented in Figure~\ref{fig:dataset1}.
\end{enumerate}

\section{Related Work}
Medical imaging is a critical aspect of modern healthcare, providing critical information to diagnose, monitor and treat various diseases and conditions. Despite technological advancement and the increasing demand for more effective and efficient medical imaging techniques, the IST application in medical imaging has been limited due to the need for more access to large datasets based on the reference literature~\cite{razzak2018deep, ahmad2017mallat, sahiner2019deep}. Below, we provide an overview of some of the referenced literature that has employed the IST in medical imaging~\cite{lan2018integrated, pereira2022melanoma, bargsten2021attention, khemchandani2022brain, jiang2020scattering}.

The scattering transform is a widely used signal processing technique in medical image retrieval due to its ability to provide extensive semantic representations of an image~\cite{lan2018simple}. However, the current features derived from the scattering transform only consider a single distribution of information~\cite{de2021scattering}. To address this limitation, Lan et al. (2018) proposed an integrated scattering feature that combines two types of compressed scattering data obtained from different perspectives using data concentration and canonical correlation analysis. This results in a more comprehensive representation of the original medical image, as demonstrated by experiments conducted on two benchmark medical CT image databases, where the proposed method outperformed several state-of-the-art techniques~\cite{lan2018integrated}. 
However, the proposed integrated approaches for medical image retrieval is time-consuming, primarily due to the scattering transform. Therefore, future research should aim to develop alternative transforms that provide similar high-level representations of medical images with reduced time consumption.

 Pereira et al. (2022) present a novel approach for skin lesion classification, focusing on distinguishing melanoma from nevus. Conventional medical image classification methods often rely solely on 2D RGB images. However, this work seeks to expand beyond these limitations by incorporating depth information that characterizes the skin surface rugosity obtained from light-field images. The authors employ a processing pipeline that utilizes a Morlet scattering transform and a CNN model to achieve this. Results show that the discrimination of melanoma and nevus reaches an accuracy of 94\% when using both 2D and 3D information, representing an improvement of 14.29 percentage points in sensitivity and 8.33 percentage points in specificity compared to using only 2D information. These findings demonstrate the significant advantages of incorporating depth information in skin lesion classification and highlight the potential of the proposed approach in improving conventional methods~\cite{pereira2022melanoma}. 
 
 However, the proposed methods have some limitations. For instance, melanoma discrimination, even with deep learning methods, remains a challenging problem, and current systems still need to achieve satisfactory sensitivity performance. Therefore, pursuing a solution to identify melanoma automatically continues to be a challenge, and current algorithms cannot achieve the desired level of sensitivity.

 Intracoronary imaging modalities like Intravascular Ultrasound (IVUS) are crucial for improving the results of percutaneous coronary interventions, and efficient extraction of essential vessel metrics such as lumen diameter, vessel wall thickness, and plaque burden through automatic segmentation of IVUS images can improve clinical workflow~\cite{nguyen2020contemporary}. However, clinical datasets are often small, leading to extracting image features that could be more meaningful and improve performance. For instance, Bargsten et al. (2021) conducted a study to explore the potential of integrating scattering transformations into Convolutional Neural Networks (CNNs) to segment small IVUS datasets. The authors proposed a novel network module that leveraged the features of a scattering transform for an attention mechanism and observed that this approach leads to improved results for calcium segmentation, lumen, and vessel wall segmentation compared to state-of-the-art data-driven methods~\cite{bargsten2021attention}.
 
While the study provides valuable insights into the potential of incorporating scattering transformations into CNNs for IVUS segmentation, it is limited to this specific application. The generalizability of these findings to other types of medical imaging or other applications remains to be determined and requires further research to be fully understood. Furthermore, it is essential to determine the conditions under which scattering transformations can be most effectively utilized to improve medical image segmentation.



Khemchandani et al. (2022) present a novel framework for brain tumor segmentation and identification using MR images. The framework combines the Density-Based Spatial Clustering of Applications with Noise (DBSCAN) algorithm for image segmentation, the Scatter Local Neighborhood Structure (SLNS) descriptor for feature extraction, and the Particle Imperialist Deep Convolutional Neural Network (PI-Deep CNN) for tumor-level classification. The proposed method showed promising results with a maximum accuracy of 0.965 on the BRATS database~\cite{khemchandani2022brain}. 
One limitation of the study is that it was only tested on a single database (BRATS), and the performance of the proposed method on other databases or real-world scenarios is still being determined. Further validation on a more extensive and diverse set of MR images is needed to establish the robustness and generalizability of the proposed method. Another area for improvement is that the proposed method can only classify the tumor into four levels; further research is needed to develop a more fine-grained classification system to provide more detailed information for clinical diagnosis and treatment planning. 

Jiang et al. (2020) present a study that utilizes scattering transform-based features for detecting epilepsy in EEG signals. The authors apply the scattering transform to extract features, namely fuzzy entropy and log energy entropy, which are inputs for five different classifiers. The results show that the proposed approach outperforms state-of-the-art techniques with high classification accuracy (99.75\% average accuracy and 0.99 Matthews Correlation Coefficient)~\cite{jiang2020scattering}. However, the study has limitations, such as the limited number of classifiers used for evaluation and the need for further validation with more diverse and extensive datasets. Additionally, the tradeoff between classification accuracy and computational time complexity achieved in this study may require further optimization for practical clinical use.

The study by Omer (2023) presents a method for lung cancer detection using Wavelet Scattering Transform (WST) and artificial intelligence techniques. The proposed method involves the representation of image features through WST and the classification of images using three machine learning algorithms: support vector machine (SVM), kernel nearest neighbor (KNN), and random forest (RF). The results indicate high accuracy in detecting the type of cancer, with the highest accuracy of 99.90\% achieved using the WST + RF network. However, the limitations of this study include its narrow focus on CT lung images and lack of evaluation of other types of lung cancer or imaging modalities. Additionally, the potential drawbacks of using WST, such as computational complexity and overfitting, still need to be addressed. Further validation of a larger and more diverse data set is necessary to establish the generalizability of the proposed method~\cite{omer2023lung}.

Abdullah et al. (2022) presented a hybrid pipeline for screening Covid-19 using chest X-rays (CXRs). The pipeline consisted of two modules, the first being a traditional Convolutional Neural Network (convnet) that generated masks of the lungs overlapping the heart and large vessels. The second module was a hybrid convnet that preprocessed the CXRs and corresponding lung masks using the Wavelet Scattering Transform. The resulting feature maps were then processed through an Attention block and a cascade of Separable Atrous Multiscale Convolutional Residual blocks to produce a classification as either Covid or non-Covid. The first module was trained on a public dataset of 6,395 CXRs with radiologist-annotated lung contours, while the second module was trained on 2,362 non-Covid and 1,435 Covid CXRs. The six distinct cross-validation models were combined into an ensemble model to classify the CXR images in the test set. The results showed that the hybrid pipeline generated high-resolution heat maps that identified the affected lung regions and enabled rapid Covid vs. non-Covid classification of CXRs~\cite{chen2023dual}.

Mohan et al. (2022) conducted a study to evaluate the performance of various classification algorithms for the classification of regions in Magnetic Resonant Images (MRI) of human brains affected by brain tumors. The study aimed to determine the best algorithm based on accuracy and dice score, which are commonly used metrics in this field. The study results indicate that the Fuzzy C Means, K Means Clustering, and Graph Cut algorithms improved accuracy, while the Deep LSTM model with N4ITK and Gaussian filters had a higher dice score. This study provides valuable information on the performance of different algorithms in detecting brain tumors in MRI images and contributes to advancing the field~\cite{mohan2022performance}.

Agboola and Zaccheus (2023) studied the application of the Invariant Scattering Convolution Network (Wavelet Scattering Network) in detecting glaucoma from retinal fundus images. They aimed to evaluate the feasibility of using automatically extracted features from retinal fundus images for glaucoma diagnosis and assess the impact of wavelet scattering network parameter settings and image type on detection accuracy. The proposed method departs from traditional approaches, which rely on the wavelet transform applied to preprocessed images and the extraction of handcrafted features. Instead, the authors fed the RIM-ONE DL image dataset into a wavelet scattering network, which performed a stage-wise decomposition process known as wavelet scattering and automatically learned features from the images. The learned features were then utilized to construct simple and computationally efficient classification algorithms. The study's results indicate the potential of a wavelet image scattering-based approach for glaucoma detection~\cite{agboola2023wavelet}.

Gaudio et al. (2023) presented DeepFixCX, a novel solution that addresses the challenge of balancing explainability and privacy in medical image analysis. This deep learning-based image compression algorithm removes or obscures spatial and edge information in an image while providing explainable post-hoc explanations of spatial and edge bias. The authors demonstrated the efficiency of DeepFixCX, which enables low-memory MLP classifiers for vision data. The algorithm was shown to improve the predictive classification performance of a Deep Neural Network (DNN) in various medical image analysis tasks such as glaucoma and cervix type detection and multi-label chest X-ray classification. These results highlight the potential of DeepFixCX as a solution to the trade-off between explainability and privacy in medical image analysis~\cite{gaudio2023deepfixcx}.

Adel et al. (2017) proposed a new method for classifying neurodegenerative brain diseases through the use of MRI scans. The authors recognized that relevant information for classification is present in the form of complicated multivariate patterns, and thus proposed using a three-dimensional scattering transform, which resembles a Deep Convolutional Neural Network (DCNN) but lacks learnable parameters. The transform linearizes diffeomorphisms in MRI scans, making it easier to separate different disease states using a linear classifier. The authors showed that scattering representations were highly effective for disease classification in experiments on Alzheimer's disease brain morphometry and HIV-related white matter microstructural damage. A visualization method was also presented to highlight areas that provide evidence for or against a certain class on an individual and group level. The results of the semi-supervised learning for the classification of progressive versus stable mild cognitive impairment reached an accuracy of 82.7\%~\cite{adel20173d}. This study highlights the potential of using a three-dimensional scattering transform for disease classification in neuroimaging and calls for further research in this field.

A comprehensive summary of some of the reference literature, their findings, and limitations is compiled in Table~\ref{tab:1}.
\footnotesize
\begin{longtable}
{@{}p{.13\linewidth}p{.2\linewidth}p{.14\linewidth}p{.12\linewidth}p{.12\linewidth}p{.16\linewidth}@{}}
\caption{Referenced literature that considered IST-based approaches in medical imaging.}\label{tab:1}\\
\toprule
  
       Reference&	Contributions&	Algorithm&	Performance&	Advantages&	Limitations  \\\midrule\endfirsthead
       \caption{\emph{Cont.}}\\\toprule
       Reference&	Contributions&	Algorithm&	Performance&	Advantages&	Limitations\\\midrule
       \endhead
       Lan et al. (2018)&	Proposed an integrated scattering feature that combines two types of compressed scattering data from different perspectives&	Data concentration and canonical correlation analysis&	Outperformed several state-of-the-art techniques on two benchmark medical CT image databases&	Provides a more comprehensive representation of the original medical image&	Time-consuming\\
        Pereira et al. (2022)&	Presented a novel approach for skin lesion classification that incorporates depth information obtained from light-field images&	Morlet scattering transform and a CNN&	Discrimination of melanoma and nevus reached an accuracy of 94\% when using both 2D and 3D information&	Significant improvement in sensitivity and specificity compared to using only 2D information&	Limited sensitivity performance, melanoma discrimination remains a challenging problem\\
     Bargsten et al. (2021)&	Explored the potential of integrating scattering transformations into Convolutional Neural Networks (CNNs) to segment small IVUS datasets. Proposed a novel network module that leveraged the features of a scattering transform for an attention mechanism&	Scattering Transform and CNN&	Improved results for calcium segmentation, lumen, and vessel wall segmentation compared to state-of-the-art data-driven methods	&Improved performance for IVUS segmentation.	&Limited to the specific application of IVUS segmentation. The generalizability of the findings remains to be determined.\\
     Khemchandani 
     et al. (2022)&
     Proposed a novel framework for brain tumor segmentation and identification using MR images& DBSCAN for segmentation, SLNS for feature extraction, PI-Deep CNN for classification& Maximum accuracy of 0.965 on the BRATS database& Promising results& Limited to a single database and four-level classification system\\

     Jiang et al. (2020)& Utilized scattering transform-based features for detecting epilepsy in EEG signals& Scattering transform for feature extraction, five classifiers for evaluation& High classification accuracy (99.75\% average accuracy)&Outperformed state-of-the-art techniques& Limited number of classifiers and need for further validation with diverse datasets\\
     Omer (2023)& Proposed a method for lung cancer detection using Wavelet Scattering Transform and AI techniques& WST for feature representation, SVM, KNN, and RF for classification& High accuracy in detecting type of cancer (99.90\% with WST + RF)& High accuracy in detecting cancer& Narrow focus on CT lung images and computational complexity of WST\\
     Abdullah et al. (2022)& Presented a hybrid pipeline for screening Covid-19 using chest X-rays (CXRs)& Density-Based Spatial Clustering of Applications with Noise (DBSCAN), Scatter Local Neighborhood Structure (SLNS), Particle Imperialist Deep Convolutional Neural Network (PI-Deep CNN)& High-resolution heat maps identified affected lung regions, rapid Covid vs. non-Covid classification&High accuracy, rapid classification&Only tested on a single database (BRATS), performance on other databases and real-world scenarios still being determined\\
     Mohan et al. (2022)& Evaluated the performance of various classification algorithms for the classification of regions in Magnetic Resonant Images (MRI) of human brains affected by brain tumors&Fuzzy C Means, K Means Clustering, Graph Cut, Deep LSTM model with N4ITK and Gaussian filters& Improved accuracy, higher dice score& Valuable information on algorithm performance, contribution to advancing the field& Limited to MRI images of human brains affected by brain tumors\\
     Agboola and Zaccheus (2023)& Studied the application of the Invariant Scattering Convolution Network (Wavelet Scattering Network) in detecting glaucoma from retinal fundus images& Wavelet Scattering Network& Potential of a wavelet image scattering-based approach for glaucoma detection& Departs from traditional approaches, automatically learns features& Limited to retinal fundus images for glaucoma diagnosis, impact of wavelet scattering network parameter settings and image type on detection accuracy still needs further evaluation\\
     Gaudio et al. (2023)& Presented DeepFixCX, a deep learning-based image compression algorithm for balancing explainability and privacy in medical image analysis& DeepFixCX& Improves predictive classification performance of a Deep Neural Network (DNN) in various medical image analysis tasks& Efficient and nimble, solution to the trade-off between explainability and privacy in medical image analysis& Further validation and optimization necessary for practical clinical use\\
     Adel et al. (2017)&Proposed a new method for classifying neurodegenerative brain diseases through the use of magnetic resonance imaging (MRI) scans&Three-dimensional scattering transform& Highly effective for disease classification, accuracy of 82.7\% in semi-supervised learning for the classification of progressive versus stable mild cognitive impairment& Linearizes diffeomorphisms in MRI scans, separates different disease states& Limited to MRI scans for neurodegenerative brain diseases, further research needed\\
    \bottomrule
    \label{ref:big table}
\end{longtable}
\normalsize
\section{Discussions and Future Research Directions}

The IST has significant implications for medical imaging, both in terms of research and practical applications. Its ability to provide robust, informative, and invariant features can enhance various medical imaging tasks and improve outcomes. Here are some ways that IST might shape the future of medical imaging: 
\begin{itemize}
    \item \textbf{Improved performance in medical imaging tasks:} By providing stable and invariant features, IST can improve the performance of machine learning models in tasks like segmentation, classification, and registration. This improvement can lead to more accurate and timely disease detection and diagnosis, better treatment planning, and enhanced disease progression or treatment response monitoring. 

\item \textbf{Enhanced interpretability:} Unlike deep learning-based approaches, IST features are handcrafted and mathematically defined, which may lead to more interpretable results. This interpretability can help clinicians better understand the underlying relationships between image features and the clinical outcome, potentially leading to increased trust and adoption of AI-based techniques in medical imaging. 
\item \textbf{Complementing deep learning approaches:} IST can be combined with deep learning techniques like CNNs to leverage the strengths of both handcrafted and learned features. For example, IST features can be used as input to deep learning models, providing invariant and informative features that can improve the performance and generalization of these models in medical imaging tasks. 

\item \textbf{Robustness in challenging scenarios:} IST features are stable and robust to noise and artifacts, making them particularly suited for medical image analysis, where images often suffer from various sources of noise and distortion. This robustness can improve performance in challenging imaging scenarios, such as low-dose CT or low-field MRI. 

\item \textbf{Adaptability to different imaging modalities:} IST can be adapted to various imaging modalities, enabling the extraction of complementary features from different medical images. By combining features from different modalities, IST can improve the performance of tasks like image segmentation, classification, or detection, leveraging the complementary information provided by each modality. 

\item \textbf{Facilitating multi-modal fusion and registration:} The invariance properties of IST features make them suitable for image registration tasks and multi-modal fusion. By matching IST features between different images, it becomes possible to align and register images from different modalities, patients, or time points, facilitating more accurate analysis and comparison of medical images. 

As IST continues to be explored and developed, it has the potential to shape the future of medical imaging by contributing to improved performance in various tasks, enhancing interpretability, and complementing deep learning approaches. By addressing the challenges and limitations of IST, researchers can unlock its full potential, leading to significant advancements in medical imaging and ultimately improving patient care and outcomes.


   \item \textbf{Automated parameter tuning and selection:} Developing methods for automatically selecting and tuning IST parameters can facilitate optimal performance for specific tasks or datasets. The benefits of this direction include improved performance, reduced manual effort, and better adaptability to different imaging modalities. However, the limitations may include the need for efficient optimization algorithms and the risk of overfitting due to excessive tuning. 


\item \textbf{Multi-modal fusion and registration:} Extending IST to facilitate multi-modal fusion and registration can improve the performance of tasks like image segmentation, classification, or detection by leveraging the complementary information provided by each modality. The benefits include better alignment and integration of different imaging modalities, leading to more accurate analysis and comparison of medical images. Limitations may include the need for robust registration algorithms and challenges in handling varying image characteristics across modalities. 

\item \textbf{Interpretability and visualization:} Developing methods for interpreting and visualizing IST features can help clinicians better understand the relationships between image features and clinical outcomes. The benefits of this direction include increased trust and adoption of AI-based techniques in medical imaging. The limitations may involve the complexity of interpreting multi-scale and multi-orientation features and developing effective visualization techniques.

\item \textbf{Adapting IST for specific medical imaging tasks:} Customizing IST for specific medical imaging tasks, such as detecting certain diseases or identifying specific anatomical structures, can improve performance in those tasks. The benefits include task-specific optimizations, better feature extraction, and improved clinical utility. The limitations might involve the need for extensive domain expertise and the risk of overfitting due to excessive customization. 

\item \textbf{Improving computational efficiency:} Developing strategies to improve the computational efficiency of IST can make it more suitable for real-time applications or when dealing with large medical image datasets. The benefits include faster processing times and broader applicability in clinical settings. The limitations may involve the trade-off between computational efficiency, feature quality, or expressiveness. 
\end{itemize}
These potential future research directions can further enhance the utility and applicability of IST in medical imaging by addressing its current limitations and challenges. However, each direction comes with its own potential benefits and limitations, which must be carefully considered when pursuing these research avenues. Ultimately, exploring these future directions can lead to significant advancements in medical imaging, contributing to improved patient care and outcomes. 

\begin{figure}[H]
    \centering
    \rotatebox{270}{\includegraphics[width=1.38\linewidth]{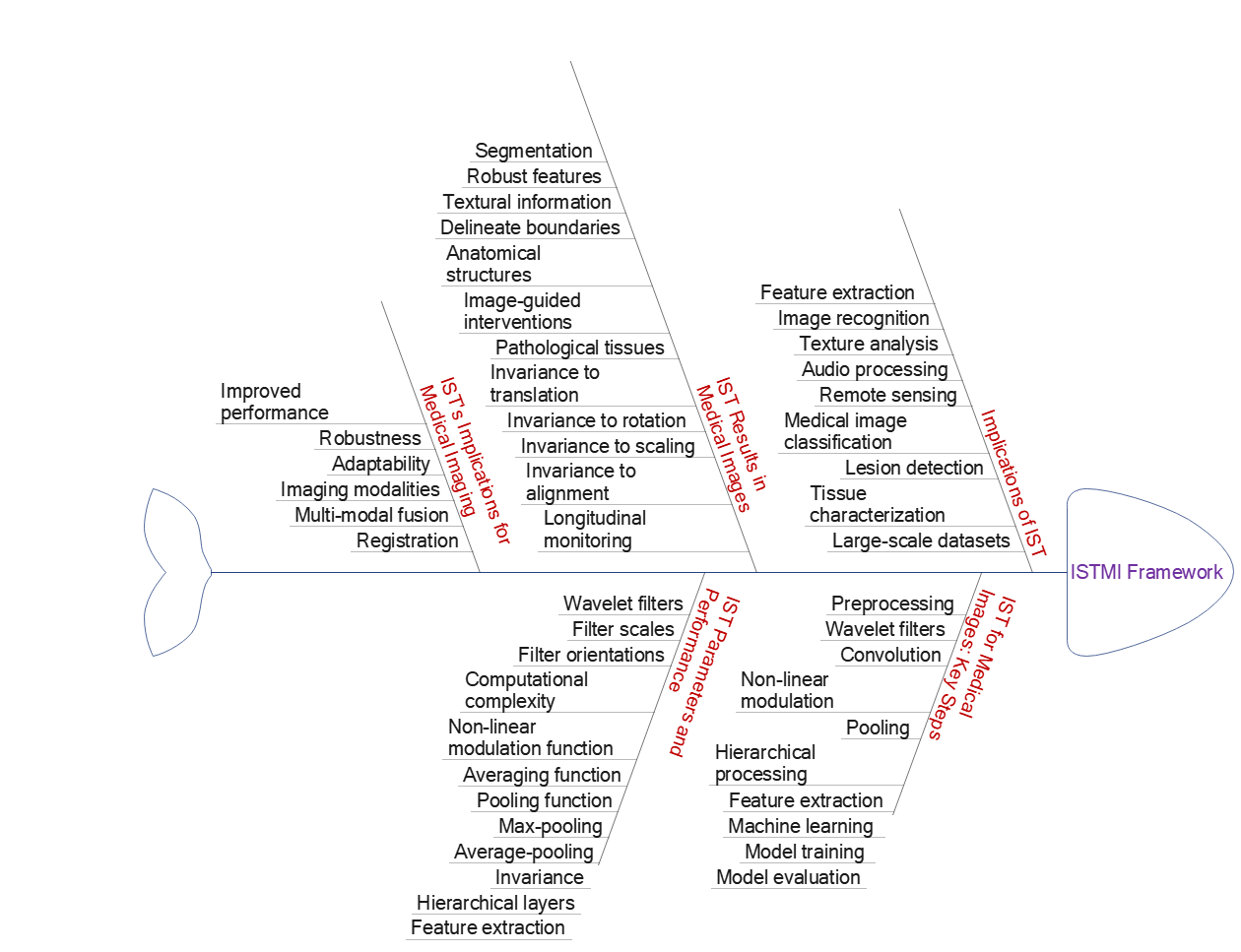}}
    \caption{ISTMI framework based on the referenced literature}
    \label{fig:fishbone}
\end{figure}

Figure~\ref{fig:fishbone} presents the fishbone framework for Invariant Scattering Transform for Medical Imaging (ISTMI) based on the relevant literature. The framework demonstrates that IST has significant implications across various domains, including feature extraction, image recognition, texture analysis, audio processing, remote sensing, and medical image classification. IST application in medical imaging entails a series of crucial steps, such as preprocessing, wavelet filters, convolution, non-linear modulation, pooling, hierarchical processing, feature extraction, machine learning, model training, and model evaluation. IST results in medical images offer numerous benefits, such as segmentation, robust features, textural information, boundary delineation, anatomical structures, image-guided interventions, pathological tissue analysis, translation, rotation, scaling, alignment invariance, and longitudinal monitoring. The parameters and performance of IST depend on several factors, such as wavelet filters, filter scales, filter orientations, computational complexity, non-linear modulation function, averaging function, pooling function (max-pooling and average-pooling), invariance, hierarchical layers, and feature extraction. IST has implications for medical imaging that include enhanced performance, robustness, adaptability, imaging modalities, multi-modal fusion, and registration, which have the potential to revolutionize medical imaging and improve patient outcomes.

\section{Conclusion}
In conclusion, the Invariant Scattering Transform (IST) presents a promising approach for medical image analysis, offering robust, invariant, and informative features that can potentially improve diagnosis, treatment planning, and monitoring. Integrating IST with machine learning algorithms can enhance medical image analysis and ultimately improve patient outcomes. IST's success in various domains underscores its potential in medical imaging applications, with significant implications for disease detection, diagnosis, and treatment planning. Although IST encounters challenges such as parameter tuning and interpretability, exploring research directions such as automated parameter tuning, combining IST with deep learning, interpretability and visualization, and task-specific adaptations can result in significant advancements in medical imaging. Overall, IST plays a crucial role in bridging the gap between traditional signal processing techniques and modern deep learning approaches, thereby contributing to advancing and improving medical imaging analysis and applications.

\bibliographystyle{unsrt}  
\bibliography{main}

\end{document}